\documentclass[pre,floatfix,a4paper,nofootinbib,amssymb,amsmath,showpacs,superscriptaddress]{revtex4}
\usepackage{graphicx}

\newcommand {\be} {\begin{equation}}
\newcommand {\ee} {\end{equation}}
\newcommand {\Be}{\begin{eqnarray*}}
\newcommand {\Ee} {\end{eqnarray*}}
\newcommand {\bey} {\begin{eqnarray}}
\newcommand {\eey} {\end{eqnarray}}
\newcommand{\bit}{\begin{itemize}}      
\newcommand{\eit}{\end{itemize}}
\newcommand{\bfl}{\begin{flusleft}}
\newcommand{\efl}{\end{flusleft}}
\newcommand{\bfr}{\begin{flushright}}
\newcommand{\ec}{\end{center}}
\newcommand{\ben}{\begin{enumerate}}    
\newcommand{\een}{\end{enumerate}}

\newcommand{\comment}[1]{}

\begin{document} 

\title{Stability of splay states in globally coupled rotators}

\author{Massimo Calamai}
\email{massimo.calamai@gmail.com}
\affiliation{Istituto dei Sistemi Complessi, CNR, via Madonna del Piano 10, I-50019 Sesto Fiorentino, Italy}
\affiliation{INFN Sez. Firenze, via Sansone, 1 - I-50019 Sesto Fiorentino, Italy}
\author{Antonio Politi}
\email{antonio.politi@cnr.it}
\affiliation{Istituto dei Sistemi Complessi, CNR, via Madonna del Piano 10, I-50019 Sesto Fiorentino, Italy}
\affiliation{Centro Interdipartimentale per lo Studio delle Dinamiche
Complesse, via Sansone, 1 - I-50019 Sesto Fiorentino, Italy}
\author{Alessandro Torcini}
\email{alessandro.torcini@cnr.it}
\affiliation{Istituto dei Sistemi Complessi, CNR, via Madonna del Piano 10, I-50019 Sesto Fiorentino, Italy}
\affiliation{INFN Sez. Firenze, via Sansone, 1 - I-50019 Sesto Fiorentino, Italy}
\affiliation{Centro Interdipartimentale per lo Studio delle Dinamiche
Complesse, via Sansone, 1 - I-50019 Sesto Fiorentino, Italy}

\begin{abstract}
The stability of dynamical states characterized by a uniform firing rate
({\it splay states}) is analyzed in a network of $N$ globally pulse-coupled
rotators (neurons) subject to a generic velocity field. In particular, we
analyse short-wavelength modes that were known to be marginally stable in
the infinite $N$ limit and show that the corresponding Floquet exponent scale
as $1/N^2$. Moreover, we find that the sign, and thereby the stability, of this
spectral component is determined by the sign of the average derivative of the
velocity field. For leaky-integrate-and-fire neurons, an analytic expression
for the whole spectrum is obtained. In the intermediate case of continuous
velocity fields, the Floquet exponents scale faster than $1/N^2$
(namely, as $1/N^4$) and we even find strictly neutral directions in a 
wider class than the sinusoidal velocity
fields considered by Watanabe and Strogatz in {\it Physica D 74 (1994) 197-253}.
\end{abstract}
   

\maketitle


\section{Introduction}\label{one}

Understanding the dynamical behaviour of highly interconnected systems is of
primary importance for neural dynamics \cite{amit}, metabolic cycles \cite{fell},
cold atoms \cite{carl}, and synchronization in general oscillators \cite{kura}. A
wide variety of interesting phenomena has been discovered, but a detailed
understanding is often lacking, to the extent that even the stability
properties of stationary states in globally coupled oscillators have not been
fully clarified. 

In this paper we study an ensemble of $N$ identical rotators, i.e. dynamical
systems characterized by a single dynamical variable, the ``phase" $x$. This
includes neural models of leaky-integrate-and-fire (LIF) type, since the
variable $x$ (the membrane potential) can be interpreted as a phase.
This is done by identifying the maximum value of the potential (the spiking
threshold, that, without loss of generality, we assume to be equal to 1) with
the minimum (the resetting value assumed to be equal to 0 - see the next
section for further details), as if they corresponded to the angles $2\pi$ and
0. More precisely, we investigate the stability of {\it splay states}
\cite{nicols}. In a splay state all rotators follow the same periodic dynamics
$x(t)$ ($x(t+T)=x(t)$) but different time shifts that are evenly distributed
(and take all multiples of $T/N$, modulus $T$). Splay states have been observed
experimentally in multimode laser systems \cite{wiese90} and electronic
circuits \cite{ashw90}. Numerical and theoretical analyses have been performed
in Josephson junction arrays \cite{nicols}, globally coupled Ginzburg-Landau
equations \cite{hakim92}, globally coupled laser models \cite{rappel94}, and
pulse-coupled neuronal networks \cite{abbott}. In the context of neuronal
networks, splay states have been also recently investigated in systems with
dynamic synapses \cite{bressloff99} and in realistic neuronal
models \cite{brunel_hansel06}.

The first detailed stability analysis of LIF neurons was performed by developing
a mean-field approach that is based on the introduction of the probability
distribution of the neuron phases \cite{abbott,vvres}. The method is expected to
work in infinite systems. More recently, another approach has been implemented
\cite{zillmer2}, which is based on the linearization of a suitable Poincar\'e map
and works for any number of oscillators. As a result, it has been discovered
that the spectrum of Floquet exponents is composed of two components: (i) the
growth rate of ``long--wavelength" perturbations - perfectly identified also
with the method described in Ref.~\cite{abbott}; (ii) the growth rate of
``short--wavelength" (SW) perturbations that cannot be characterized with
methods
that involve a coarse-graining over small scales. As discussed in \cite{zillmer2},
the latter component plays a crucial role when the width of the transmitted
pulses is comparable to or smaller than $T/N$, since it may give rise to
instabilities of otherwise stable patterns. The same analysis has also
revealed that for finite pulse-widths, SW are marginally
stable in the infinite $N$-limit. It is therefore important to investigate more
thoroughly finite systems, because it is still unclear whether and when they
are stable.

Here we address precisely this question, by first implementing a perturbative
technique in the standard LIF model and by then numerically investigating the
behaviour of a more general class of rotators, characterized by a nonlinear
velocity field $F(x) = \dot x$. All of our results indicate that the SW 
component scales as $1/N^2$ if and only if $F(1)\ne F(0)$. Moreover, we
systematically find that the SW component is stable (resp. unstable) if
$F(1)<F(0)$ (resp. $F(1)>F(0)$. Since $\Delta F=F(1)-F(0)$ is, by definition, the
average derivative of $F$, the two classes of systems will be identified as
decreasing and increasing fields, respectively.

At the boundary between these two classes of fields, continuous velocity fields
($F(1)=F(0)$) turn out to exhibit a faster scaling to zero of the Floquet
SW spectrum. In the case of analytic functions, many exponents appear even to
be numerically indistinguishable from zero. This scenario is coherent with, and
in some sense extends, the theorem proved in \cite{watanabe}, where it has been
shown that in the presence of a sinusoidal field
$F = a(t) + sin(2\pi x + \alpha)$, one should expect $N-3$ zero exponents for
any dependence of $a(t)$. 

The paper is organized as follows. In section \ref{two} we introduce the model
and the event-driven map that is used to carry out the stability analysis.
In Sec.~\ref{three} we derive analytical perturbative expressions for the
Floquet spectrum in the case of LIF neurons. The results are compared with
the numerical solution of the exact equation. In Sec.~\ref{four} we numerically
analyse several examples of velocity fields to test the validity of the
above mentioned conjectures. Finally, in Sec.~\ref{five}, we summarize the
main results and the open problems.


\section{The model}\label{two}

We consider a network of $N$ identical neurons (rotators) coupled via a
mean-field term. The dynamics of the $i$-th neuron writes as
\begin{equation}\label{eq:x1}
  \dot{x}_{i}= F(x_{i})+gE(t)\, \\
\end{equation}
where $x_i$ represents the membrane potential, $E(t)$ is the ``mean" forcing
field, and $g$ is the coupling constant, the analysis will be limited to the
excitatory case, i.e. $g > 0$. When the membrane potential reaches the
threshold value $x_i=1$, a spike is sent to all neurons (see below for the
connection between single spikes and the global forcing field $E$) and it is
reset to $x_i=0$. The resetting procedure is an approximate way to describe
the discharge mechanism operating in real neurons. The function $F(x)$
is assumed to be everywhere positive (thus ensuring that the neuron is
repetitively firing, i.e. it is supra-threshold). For $F(x)=a-x$, the model
reduces to the well known case of leaky integrate-and-fire (LIF) neurons.
The field $E$ is the linear superposition of the pulses emitted in the past 
when the membrane potential of each single neuron has reached the threshold
value. By following Ref.~\cite{abbott}, we assume that the shape of a pulse
emitted at time $t=0$ is given by 
$E_s(t)= \frac{\alpha^2 t}{N} {\rm e}^{-\alpha t}\,$, where $1/\alpha$ 
is the pulse--width. This is equivalent to saying that the total field evolves
according to the equation
\begin{equation}\label{eq:E}
  \ddot E(t) +2\alpha\dot E(t)+\alpha^2 E(t)= 
  \frac{\alpha^2}{N}\sum_{n|t_n<t} \delta(t-t_n) \ .
\end{equation}
where the sum in the r.h.s. represents the source term due to the spikes
emitted at times $t_n<t$.

\subsection{Event-driven map}

As anticipated in the introduction, it is convenient to transform the
differential equations into a discrete-time mapping. We do so by integrating 
Eq.~(\ref{eq:E}) from time $t_n$ to time $t_{n+1}$ (where $t_n$ is the
time immediately after the $n$-th pulse has been emitted). The resulting map
reads
\begin{subequations}\label{eq:map}
\begin{gather}
  E(n+1)=E(n) {\rm e}^{-\alpha \tau(n)}+Q(n)\tau(n) 
{\rm e}^{-\alpha \tau(n)}\\
  Q(n+1)=Q(n)e^{-\alpha \tau(n)}+\frac{\alpha^2}{N}\ ,
\end{gather}
\end{subequations}
where $\tau(n)= t_{n+1}-t_n$ is the interspike time interval
and, for the sake of simplicity, we have introduced the new 
variable $Q := \alpha E+\dot E$~. 

Moreover, the differential equation (\ref{eq:x1}) can be formally integrated to
obtain,
\begin{equation}\label{eq:xgen}
  x_{j}(n+1)={\cal F}(x_j(n),E(n),Q(n),\tau(n)) 
\quad j=1,\dots,N-1 \qquad; \qquad x_m(n+1) \equiv 0
\end{equation}
where $m$ indicates the closest-to-threshold neuron at time $n$ and
the time interval $\tau(n)$ is determined by imposing the condition that $x_m$
reaches the value 1 at time $n+1$, immediately before being reset to zero.
Altogether, we have therefore transformed the initial problem into a discrete
time map for $N+1$ variables: $E$, $Q$ and $N-1$ membrane potentials (where
one degree of freedom is eliminated as a result of taking the
Poincar\'e-section). A relevant property of identical mean-field coupled 
rotators is that the ordering of the local variables is preserved by the
dynamical evolution: all neurons ``rotate" around the circle $[0,1]$ 
(1 being identified with 0) without passing each other. On the other hand,
being the neurons identical, we can change their labels as they are
indistinguishable. By following Ref.~\cite{zillmer,zillmer2}, it is convenient to
start ordering the membrane potentials from the largest to the smallest one
and then to introduce a comoving reference frame, i.e. to decrease
by 1 the label of each neuron (plus $1 \to N$) at each step of the iteration.
In this frame, the label of the closest-to-threshold neuron is always equal to
1 and the splay state is just a fixed point of the transformation. Accordingly
the linear stability analysis amounts to determining the eigenvalues of the
corresponding linearized transformation.

In order to carry out the stability analysis, it is necessary to derive
an explicit expression for ${\cal F}(x_j(n),E(n),Q(n),\tau(n))$. This is
not generally doable, but in the thermodynamic limit ($N\to\infty$) one can
exploit the smallness of $\tau \sim {\cal O}(1/N)$ and correspondingly
set up a suitable perturbative expansion. We shall see that in order to
correctly reproduce the stability of the splay state in typical cases, it is
necessary to expand the map to fourth order. In a few peculiar models, a fully
analytic calculation is possible. This is the case of LIF neurons, because of
the linear structure of the velocity field: they will be analysed in the next
section.

\section{Leaky integrate-and-fire model}\label{three}

Let us now consider the leaky integrate-and-fire case
for the supra-threshold neuron, namely $a >1$.  
In the comoving frame, Eq.~(\ref{eq:xgen}) writes as, \cite{zillmer,zillmer2}
\begin{equation}\label{eq:xx2}
  x_{j-1}(n+1)=x_j(n)e^{-\tau(n)}+1-x_1(n)e^{-\tau(n)}\ 
\qquad j=1,\dots,N-1 \enskip,
\end{equation}
with the boundary condition $x_N=0$, while the $n$-th interspike interval is
given by the self-consistent equation,
\begin{equation}\label{eq:ti2}
  \tau(n)=\ln\left[\frac{a-x_1(n)}{a+gH(n)-1}\right]\ ,
\end{equation}
where,
\begin{equation}\label{eq:F1}
  H(n)= \frac{{\rm e}^{-\tau(n)} - e^{-\alpha\tau(n)}}{\alpha-1}
     \left(E(n)+\frac{Q(n)}{\alpha-1} \right) - 
  \frac{\tau(n) e^{-\alpha\tau(n)}}{(\alpha-1)} Q(n)~.
\end{equation}
In the absence of coupling ($g=0$), ${\rm e}^\tau$ is obviously equal to the
ratio of the initial ($a-x_1$) and final ($a-1$) velocity of the first neuron,
since the dynamics reduces to a pure relaxation. 
In that case there would be no way to
determine $\tau$ as the r.h.s. would be independent of it,
but with interactions this is no  longer true.
As soon as the coupling is switched on, the velocity starts depending on the evolution of the
field $E$ in the way that it is summarized by the expression $H(n)$.

The set of Eqs.~(\ref{eq:map},\ref{eq:xx2},\ref{eq:ti2},\ref{eq:F1}) defines a 
discrete-time mapping that is perfectly equivalent to the original set of 
ordinary differential equations. It should be noticed that Eq.~(\ref{eq:F1})
is valid for any physically meaningful pulse-width value (i.e., $\alpha > 0$)
including $\alpha=1$, when there is no divergence or discontinuity.
Moreover, in the parameter region considered in this paper (i.e. $g > 0$ and
$a >1$) the logarithm in Eq.~(\ref{eq:ti2}) is well defined, since one can show
that $H(n)$ is always positive.

In this framework, the splay state reduces to a fixed point that
satisfies the following conditions,
\begin{subequations}\label{eq:ps}
\begin{gather}
  \tau(n)\equiv \frac{T}{N}\ ,\\
  E(n)\equiv\tilde E\, ,\ Q(n)\equiv\tilde Q\ ,\\
  \tilde x_{j-1}=\tilde x_j e^{-T/N}+1-\tilde x_1e^{-T/N}\ ,
\end{gather}
\end{subequations} 
where $T$ is the time elapsed between two consecutive spike emissions of
the same neuron. A simple calculation yields,
\be
\label{eq:eqstat}
  \tilde Q=\frac{\alpha^2}{N}\left(1-e^{-\alpha T/N}\right)^{-1}\,,\ 
  \tilde E=\tau \tilde Q\left(e^{\alpha T/N}-1\right)^{-1}\ .
\ee
The solution of Eq.~(\ref{eq:ps}c) involves a geometric series that, together
with the boundary condition $\tilde x_N=0\,$, leads to a transcendental equation
for the period $T\,$. This, in the large $N$ limit and at the leading order,
reduces to the following simple expression:
\begin{subequations}\label{eq:perxt}
\begin{gather}
  \tilde x_{j}=\frac{{\rm e}^T - {\rm e}^{j\,\tau}}{{\rm e}^T-1} ,\\
  T=\ln\left[\frac{aT+g}{(a-1)T+g}\right]\ .
\end{gather}
\end{subequations}
The lack of any dependence of the period from the pulse--width is due to
the fact that in the $N \to \infty$ limit the forcing
field reduces to $\tilde E = 1/T$ \cite{abbott,zillmer2}.
If we assume that $a > 1$ (which corresponds to assuming that the single neuron
is supra-threshold), we see that in the excitatory case $(g>0$) the period $T$ is
well defined only for $g < 1$ ($T\to 0$, when $g$ approaches 1), while in 
the inhibitory case ($g<0$), a meaningful solution exists for any coupling 
strength ($T\to \infty$ for $g \to -\infty$).

\subsection{Linear stability}

By linearizing Eqs.~(\ref{eq:map},\ref{eq:xx2}) around the fixed point
(\ref{eq:ps}), we obtain
\begin{subequations}\label{eq:lin1}
\begin{gather}
  \delta E(n+1)=e^{-\alpha \tau}\delta E(n)+ \tau e^{-\alpha \tau}\delta Q(n)
  -\left(\alpha\tilde E-\tilde Qe^{-\alpha \tau}\right)\delta \tau(n)\,,\\
  \delta Q(n+1)=e^{-\alpha \tau} \left[ \delta Q(n)-\alpha\tilde Q 
  \delta \tau(n) \right] \, ,\\
  \delta x_{j-1}(n+1)=e^{-\tau}[\delta x_j(n)-\delta x_1(n)]+ e^{-\tau}
  (\tilde x_1-\tilde x_j)\delta \tau(n)\,,
\end{gather}
\end{subequations}
and the expression for $\delta \tau(n)$ can be derived by linearizing
Eqs.~(\ref{eq:ti2},\ref{eq:F1}) 
\begin{equation}
\label{eq:tauder}
  \delta \tau(n) =\tau_x \delta x_1(n) +\tau_E\delta E(n)+\tau_Q\delta Q(n)\ ,
\end{equation}
where $\tau_x:=\partial \tau/\partial x_1$ and analogous definitions are
adopted for $\tau_E$ and $\tau_Q$. 

In the comoving frame, the boundary condition $x_N\equiv 0\,$ implies
$\delta x_N=0\,$. In practice, the stability problem is solved by computing
the Floquet spectrum of multipliers $\{\mu_k\}\,$, $k=1,\ldots,N+1\,$ 
corresponding to the linear evolution \eqref{eq:lin1}~. It should be stressed that 
in general, the solution can be determined only numerically.

\begin{figure}[t!]
\includegraphics[draft=false,clip=true,height=0.34\textwidth]{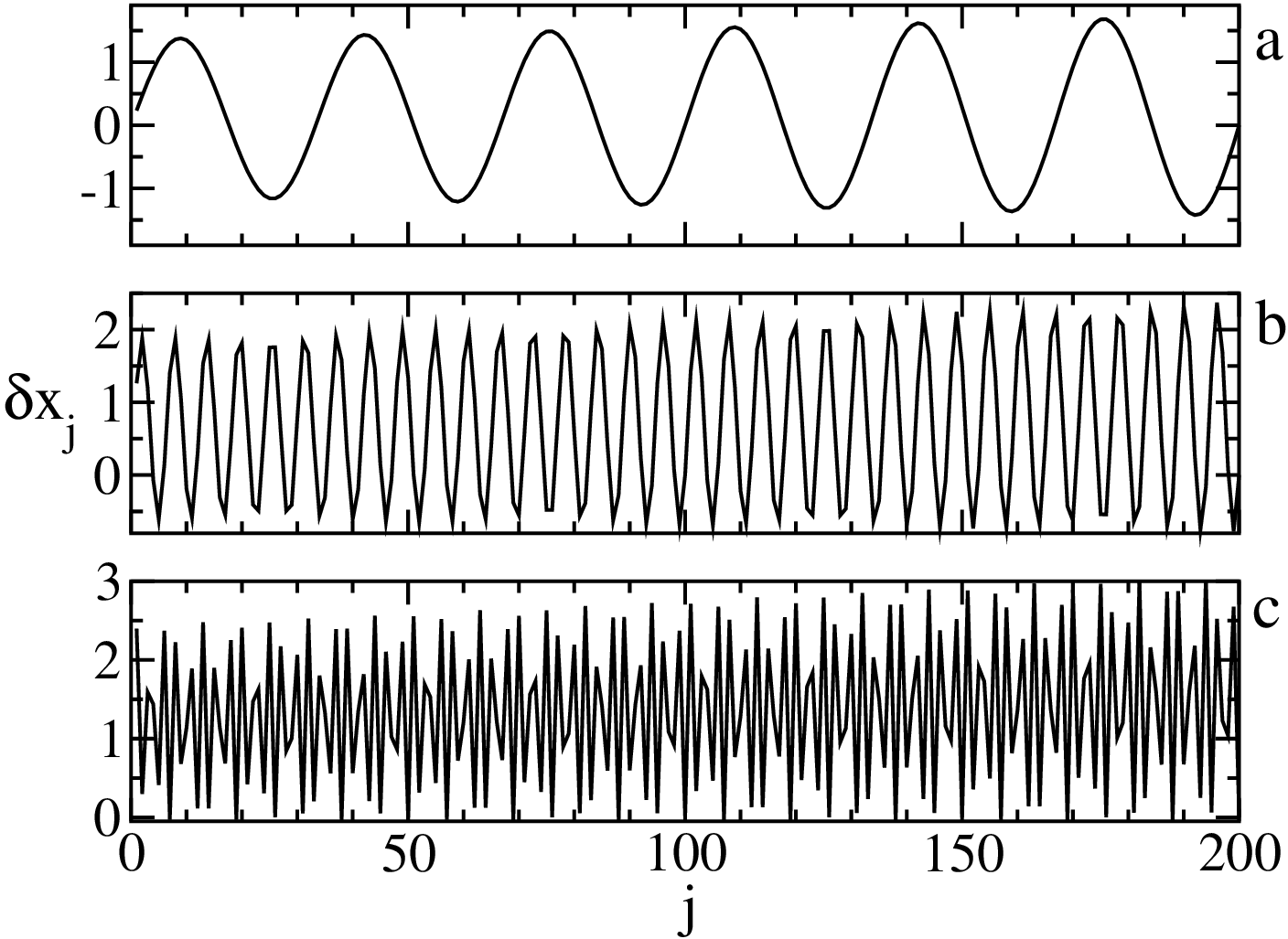}
\caption{Three instances of the real part of eigenvectors for LIF neurons for $a=3$,
$g=0.4$, $\alpha=3$, and $N=200$. From top to bottom, 
panels (a-c) correspond to $\varphi = 0.06\pi$, $0.34\pi$, and $0.78\pi$.}
\label{fig1}
\end{figure}

However, it is convenient to rewrite the Floquet multipliers as
\begin{equation}\label{eq:specdef}   
  \mu_k=e^{i\varphi_k}e^{T(\lambda_k+i\omega_k)/N}\,,\ 
\end{equation}
where $\varphi_k= \frac{2\pi k}{N}\,,\ k=1,\ldots,N-1$ and $\varphi_N =
\varphi_{N+1}=0$, while $\lambda_k$ and $\omega_k$ are the real and imaginary
parts of the Floquet exponents. 
The variable $\varphi_k$ plays the role of the wave-number in the
linear stability analysis of spatially extended systems and one can say that
$\lambda_k$ characterizes the stability of the $k$--th mode. 
Previous studies \cite{zillmer2} have shown that the spectrum can be decomposed
into two components depending on the index $k$: (i) $k \sim \mathcal{O}(1)$; (ii)
$k/N \sim \mathcal{O}(1)$. The first component corresponds to long-wavelength
perturbations that can be formally analysed by taking the continuum limit
(this was implicitely done in Ref.~\cite{abbott}); the second component
corresponds to ``high" frequency oscillations that require taking in full
account the discreteness of the ``spatial" index $j$. This is clearly
illustrated in Fig.~\ref{fig1}, where we have plotted the spatial component
$\delta x_j$ of (the real part of) three eigenvectors. The vector plotted in
panel {\it a)} corresponds to $\varphi_k = 0.06\pi$ and is indeed both rather
smooth
and close to a sinusoidal function. In the other two panels, we can see that
upon increasing the wave-number $\varphi_k$, the discontinuous structure of the
eigenvectors becomes increasingly evident.

\begin{figure}[tbp]
\includegraphics[draft=false,clip=true,height=0.34\textwidth]{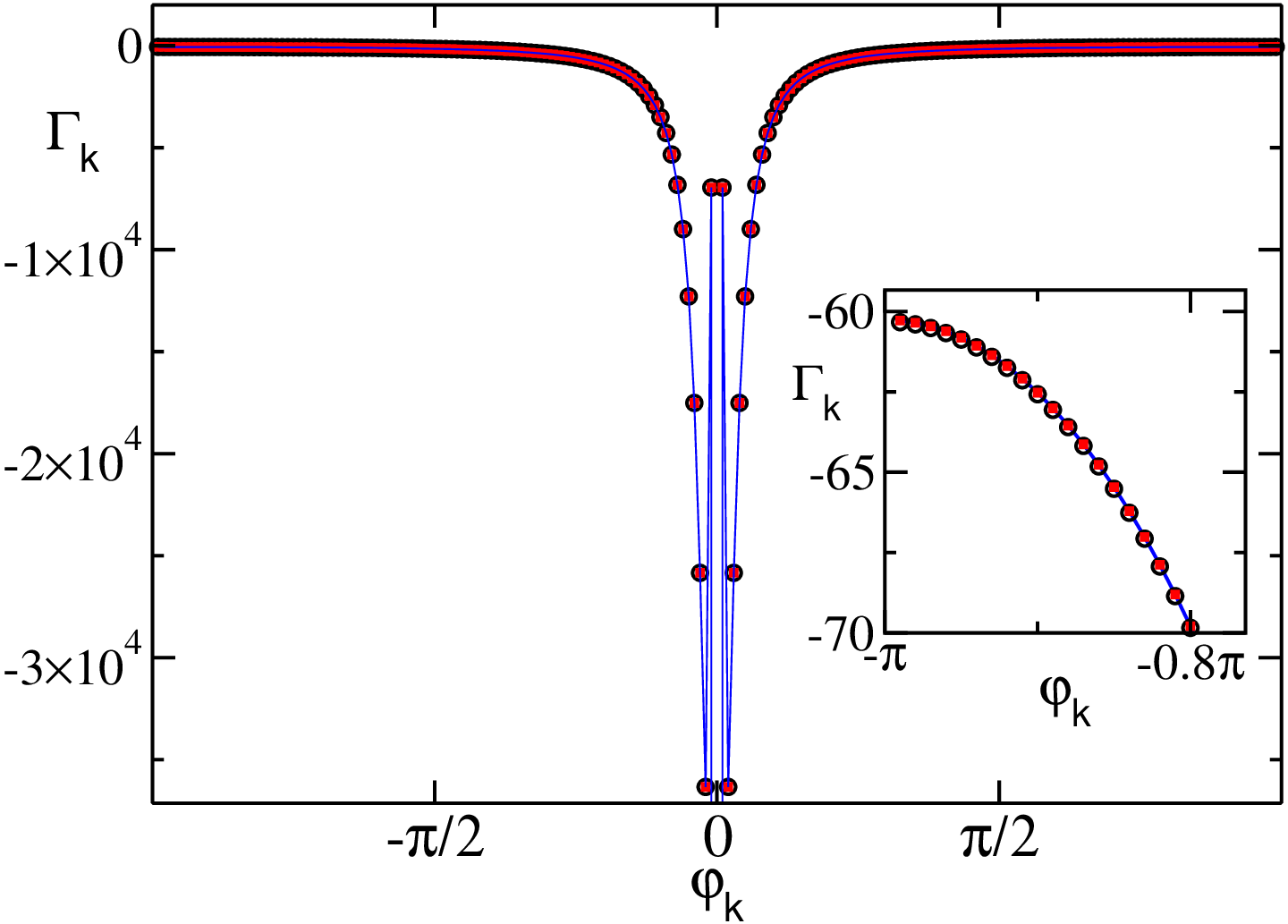}
\caption{
(Color online) Floquet exponent spectra for the LIF neurons: exact expression (\ref{equtot})
(filled red squares), perturbative expression (\ref{perturbative_spectra}) 
(blue line) and event driven map correct up to the fourth order in $\tau$
(empty black circles). The parameters are $a= 3.0$, $g=0.4$, and $\alpha=30.0$,
$N=200$.
}
\label{fig2}
\end{figure}

While the eigenvalues of the first component are of order 1, the analysis
carried out in Ref.~\cite{zillmer2} has revealed that the second component
vanishes in the $N\to\infty$ limit. Therefore it is necessary to go beyond
the zeroth order result to determine the stability of a splay state in large
but finite system. As an example, in Fig.~\ref{fig2} we show the spectrum of
the Floquet multipliers of the splay state for excitatory coupling ($g > 0$)~
and finite values of $N$~. 

\subsection{Analytical Results}

In the LIF model, many steps towards the determination of the Floquet
exponents can be performed exactly. We start by deriving expressions that are
valid for any number $N$ of neurons and eventually introduce a perturbative
approach to obtain an explicit expression in the large $N$ limit.

\subsubsection{Exact Expressions}

We start by introducing the standard Ansatz
\be
\delta E(n+1) = \mu_k \delta E(n) \qquad ; \qquad \delta Q(n+1) = \mu_k \delta Q(n)
\ee
From Eq.~(\ref{eq:lin1}a,b) 
\begin{eqnarray}
\delta Q &=& - \frac{\alpha \tilde Q}{\mu_k{\rm e}^{\alpha \tau}-1} \delta \tau \\
\delta E &=& -\left[
\frac{\alpha \tau \tilde Q}{(\mu_k{\rm e}^{\alpha \tau}-1)^2} +
\frac{\alpha \tilde E -\tilde Q +\alpha \tilde Q 
\tau}{\mu_k{\rm e}^{\alpha \tau}-1}
\right] \delta \tau
\end{eqnarray}
By combining the above equations with Eq.~\eqref{eq:tauder}, we find
\be
\label{aexp}
  x_{\tau} = \tilde x_1 - a +
\frac{g \tilde Q {\rm e}^{-(\alpha-1)\tau}}{\alpha-1} +
\frac{g \mu_k {\rm e}^{\alpha \tau}}
    {(\alpha-1)(\mu_k {\rm e}^{\alpha \tau}-1)}\left[
 \alpha \tilde E + \frac{\tilde Q}{\alpha-1}
  +  \alpha \tau \tilde Q \frac{(1-\mu_k {\rm e}^{\tau})}  
  {(\mu_k {\rm e}^{\alpha \tau}-1)} \right ]
\ee
where $x_\tau$ denotes the derivative of $x_1$ with respect to $\tau$.
From the evolution equation for $\delta x_j$, Eq.~(\ref{eq:lin1}c), and by
assuming that $\delta x_j (n+1) = \mu_k \delta x_j (n)$, we obtain
\be
\mu_k \delta x_{j-1} = {\rm e}^{-\tau}(\delta x_j-\delta x_1) +
(\tilde x_1- \tilde x_j){\rm e}^{-\tau}\delta \tau \, .
\ee
By using $\delta \tau = \delta x_1/x_\tau$ and introducing the expression
(\ref{eq:ps}c) for $\tilde x_j$, we find
\be
\mu_k \delta x_{j-1} = {\rm e}^{-\tau}\delta x_j +
 \frac{{\rm e}^{(j-1)\tau}}{{\rm e}^{T}-1} \frac{\delta x_1}{x_\tau} -
\left(
\frac{1}{x_\tau({\rm e}^T-1)} + {\rm e}^{-\tau} \right) \delta x_1 \quad .
\ee
The solution of this recursive equation reads
\be
\delta x_{j} = -\left(
\frac{1}{x_\tau({\rm e}^T-1)} + {\rm e}^{-\tau} \right)
\frac{\delta x_1}{\mu_k-{\rm e}^{-\tau}} +
\frac{{\rm e}^{j\tau}}{x_\tau({\rm e}^T-1)}\frac{\delta x_1}{\mu_k-1}
+ K \mu_k^j{\rm e}^{j\tau}  \, .
\ee
We can determine the constant $K$ by imposing that the above equation is an
identity for $j=1$. As a result,
\begin{equation}
\frac{\delta x_{j}}{\delta x_1} = \left(
\frac{1}{x_\tau({\rm e}^T-1)} + {\rm e}^{-\tau} \right)
\frac{\mu_k^{j-1}{\rm e}^{(j-1)\tau}-1}{\mu_k-{\rm e}^{-\tau}} -
\frac{{\rm e}^{j\tau}}{x_\tau({\rm e}^T-1)}\frac{\mu_k^{j-1}-1}{\mu_k-1}
+ \mu_k^{j-1}{\rm e}^{(j-1)\tau}
\end{equation}
The equation for the determinant is finally obtained by imposing
$\delta x_N=0$,
\begin{equation}
x_\tau ({\rm e}^{T}-1)\mu_k^{N-1} =
-\left( x_\tau({\rm e}^{T}-1) + {\rm e}^{\tau} \right)
\frac{{\rm e}^{\tau-T}-\mu_k^{N-1}}{1-\mu_k{\rm e}^{\tau}} +
{\rm e}^{\tau}\frac{1-\mu_k^{N-1}}{1-\mu_k}
\label{equtot}
\end{equation}
Eq.~\eqref{equtot} is an exact but implicit expression for all Floquet multipliers
that applies for a generic number $N$ of neurons.  A numerical solution of
Eq.~\eqref{equtot} reveals that, for finite $N$, excitatory coupling, and $\alpha$
smaller than the critical value $\alpha_c=\alpha_c(g,N)$ (see also \cite{vvres}), 
the splay state is strictly stable, although the maximum Floquet exponent
approaches zero for increasing $N$ \cite{zillmer2}. In fact, as shown in
Ref.~\cite{zillmer2}, in the limit $N \to \infty$, SW modes are
marginally stable, i.e. $\lambda_k \equiv \omega_k \equiv 0$. 

\subsubsection{Perturbative Expansion}

Since the Floquet exponents of the SW vectors are exactly equal to
zero in the infinite $N$-limit, it is natural to investigate the stability
of finite systems in a perturbative way. In particular, we find it convenient to
introduce the smallness parameter $\tau \simeq 1/N$. A posteriori, and in
agreement with the numerical observations in \cite{zillmer2}, it turns out that
an expansion up to second order in $\tau$ is necessary and sufficient to
correctly determine the leading contribution of the Floquet spectrum.
 
Let us start by expanding the stationary solution for $\tilde Q$ and $\tilde E$
\eqref{eq:eqstat}
\begin{subequations}
\label{eq:eqexp}
\begin{gather}
\tilde Q = \frac{\alpha}{T}\left(1 + \frac{\alpha}{2}\tau +
  \frac{\alpha^2}{12}\tau^2\right) \\
\tilde E =\frac{1}{T}\left(1 - \frac{\alpha^2}{12}\tau^2\right)
\end{gather}
\end{subequations}
Next, we express the Floquet multipliers as
\be
\label{eq:gam}
\mu_k = {\rm e}^{i\varphi_k} {\rm e}^{\Gamma_k \tau^3}  \qquad
\mu_k^N = {\rm e}^{\Gamma_k T\tau^2}  \qquad
\ee
which amounts to assume that the Floquet exponent is proportional to $\tau^2$.

By expanding $x_\tau$ and with the help of Eq.~(\ref{eq:perxt}b), we obtain
\be
x_\tau = -\frac{1+\tau-B\tau^2}{{\rm e}^T-1}
\ee
where
\be
\label{eq:defb}
B = -\frac{1}{2} + \frac{g\alpha^2}{T} \left[\frac{1}{12} +
\frac{ {\rm e}^{i\varphi_k}({\rm e}^T-1)} {({\rm e}^{i\varphi_k}-1)^2} \right]
\ee
After inserting the above expansions into Eq.~(\ref{equtot}), we obtain
\begin{equation}
(-1-\tau+B\tau^2)(\mu_k^{N-1}-\mu_k^N) =
-\tau^2 \left( B + \frac{1}{2}\right)
({\rm e}^{-T}-{\rm e}^{-i \varphi_k}) +
{\rm e}^{\tau}(1-\mu_k^{N-1}) \, .
\end{equation}
Now, with the help of Eq.~(\ref{eq:gam}) and completing the $\tau$ expansion,
\begin{equation}
\Gamma_k T  =
\left( B + \frac{1}{2}\right)
(1-{\rm e}^{-T}) \, .
\end{equation}
By finally replacing the definition \eqref{eq:defb} of $B$ into the above
equation, we obtain an explicit expression of the Floquet spectrum,
\begin{equation}
\frac{\lambda_k}{\tau^2} = \Gamma_k  = \frac{g\alpha^2}{12T^2}
({\rm e}^{T}-2 +{\rm e}^{-T})
\left[1 + \frac{6} {(\cos \varphi_k-1)}\right]
\label{perturbative_spectra}
\end{equation}
In Fig.~\ref{fig2}, Eq.~\eqref{perturbative_spectra} is compared with the numerical
but exact solution of Eq.~\eqref{equtot} for $N=200$, revealing a very good
agreement. The divergence of this expression for $\alpha \to \infty$ 
indicates that the SW modes are characterized by a different scaling behaviour
for $\delta$-like pulses. In fact, as shown in Ref.~\cite{zillmer2}, the
corresponding exponents do not scale to zero for $N\to\infty$. Moreover, it is
worth recalling that the stability of networks with strictly $\delta$-pulses
cannot be inferred by taking the limit $\alpha\to \infty$ of the exact
(non-perturbative) expressions \cite{zillmer2}.

\section{General Case}\label{four}

In the previous section we have seen that in the LIF model the SW
component of the spectrum scales as $1/N^2$ and obtained an analytic
expression for the leading term. It is natural to ask whether the observed
scaling behaviour is peculiar to this system or it is a general characteristics
of pulse coupled oscillators. For $F(x)= a + \sin(2\pi x+\alpha)$, the general
theorem proved in \cite{watanabe} tells us that the dynamics of the oscillators
is characterized precisely by $N-3$ zero exponents independently of the
behaviour of the forcing field $E$. Therefore, it is an example of perfect
neutral stability for any $N$.
For generic velocity fields, it is not possible to obtain analytic expression,
so that one has to rely on approximate expressions. Since we expect 
$\lambda_k \to 0$  for increasing $N$, it is natural to follow a perturbative
approach from the very beginning, i.e. from the definition of the event-driven
map. In \cite{zillmer2}, it has been shown that a second order expansion fails
to reproduce the stability properties of the LIF model even on a qualitative
level. In fact, the Floquet exponent of the single-step map is of order
$\tau^3$. This would naively suggest that a third order is sufficient; however,
the eigenvalue equation involves an ``integration" over $N$ steps. Therefore, it
is necessary to control the accuracy of the single iterate of the map up to
order $1/N^4$, what is incidentally guaranteed by standard integration
algorithms like fourth-order Runge-Kutta. Here he have preferred to determine
an explicit expression for the map to be thereby linearized and used to
determine the entire Floquet spectrum. The results plotted in Fig.~\ref{fig2}
confirm that upon including terms up to $O(1/N^4)$, we are able to reproduce the
expected results.

\begin{figure}[t!]
\includegraphics[draft=false,clip=true,height=0.34\textwidth]{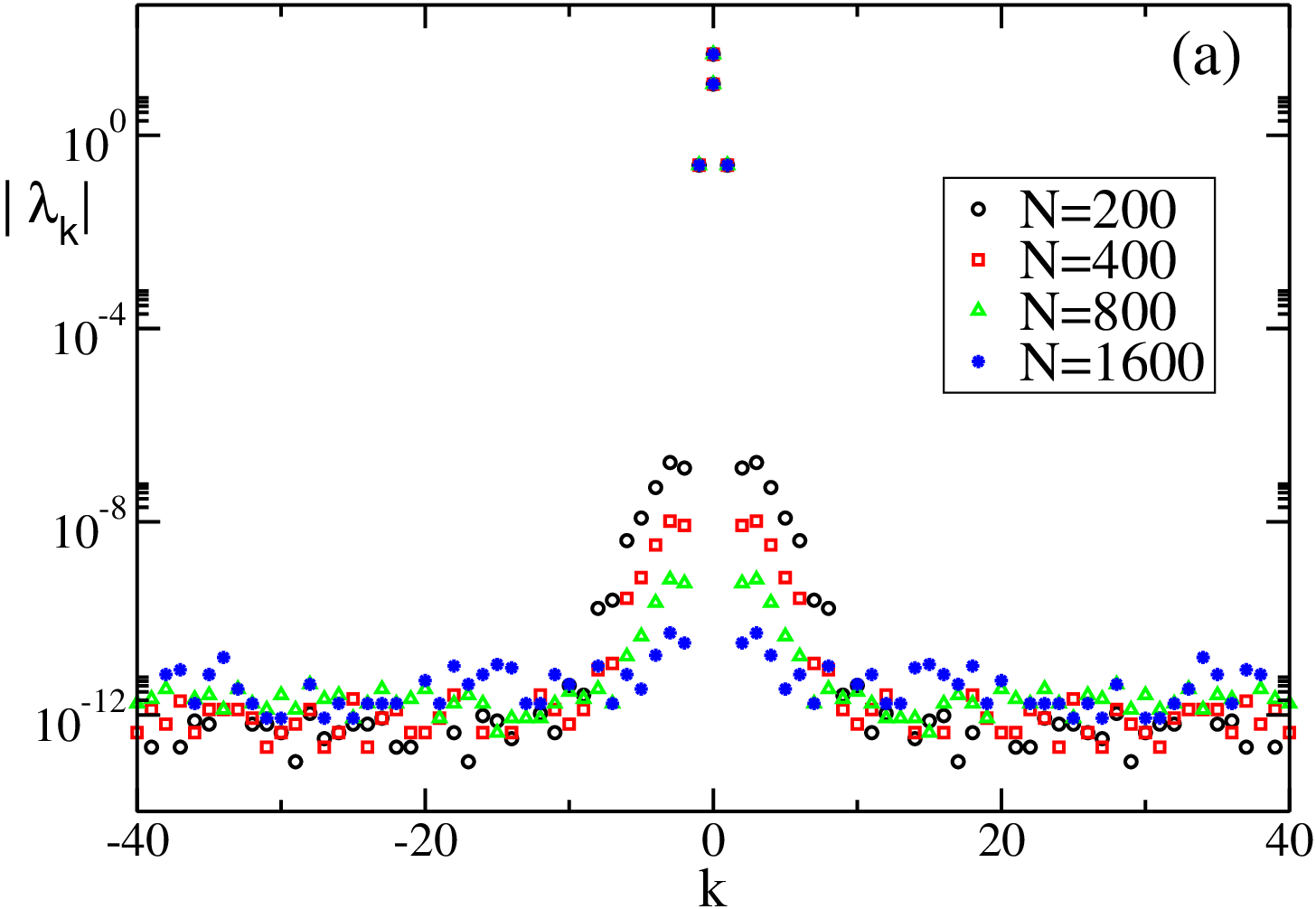}
\includegraphics[draft=false,clip=true,height=0.34\textwidth]{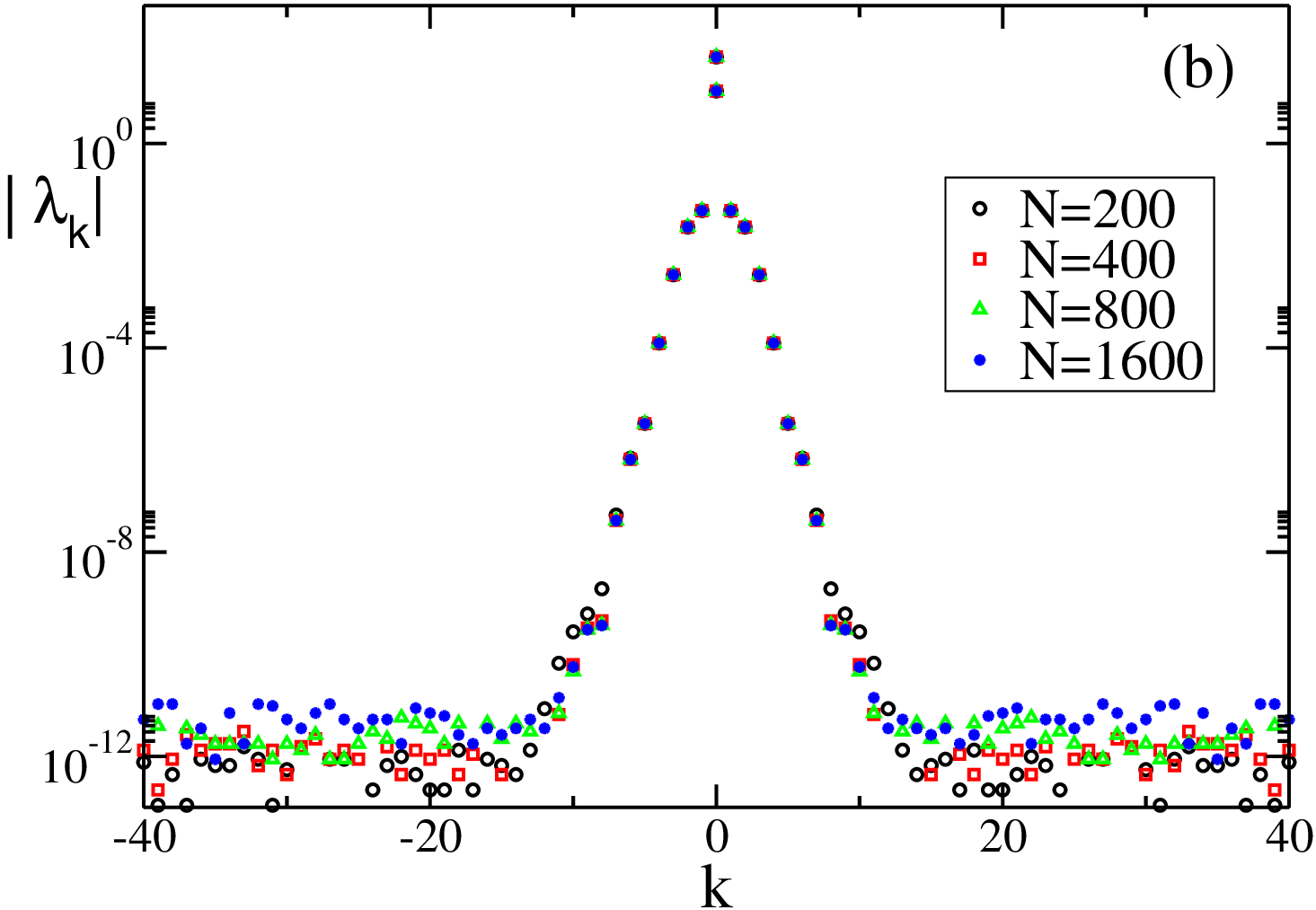}
\caption{
(Color online) Floquet spectra: (a) single harmonic; (b) three harmonics.
The parameters are $a= 3.0$, $g=0.4$, and $\alpha=30.0$.
}
\label{fig3}
\end{figure}

With the goal of identifying the typical scaling behaviour of the Floquet
spectrum, in the following we investigate various types of functions $F(x)$,
starting from smooth periodic functions. As mentioned above, if $F(x)$ contains
just the main harmonic, i.e. if $F(x) = a - \sin(2 \pi x)$, we expect $N-3$ zero
exponents ~\cite{watanabe}. In Fig.~\ref{fig3}a, we plot the spectrum for $a=3$ , $g=0.4$,
$\alpha=30$ and different numbers of rotators. The data indeed show that 4
exponents remain finite for $N \to \infty$, in agreement with the
the theoretical results~\cite{watanabe}, since in our system there are $N+1$ degrees of freedom. 
Moreover, we can see that the
vast majority of the exponents are equal to zero within numerical accuracy.
This is even beyond our expectations, because of the finite (fourth order)
accuracy of the numerical computations.

In the presence of more harmonics, there are no theoretical predictions which
can guide us. In Fig.~\ref{fig3}b, we present the results for
$F(x) = 3 - \sin(2 \pi x)/2 - 0.1 \sin(4 \pi x) - 0.01 \sin(6 \pi x)$. There
we see that there are no substantial differences from the previous case, the
main novelty being that the number of (negative) exponents which remain finite
for $N \to \infty$ is definitely larger than 4, probably 24 or 26 with our
numerical accuracy. The presence of many zero exponents is confirmed by
simulations performed for different parameter values. Whether the ``numerical
zeros" correspond to exact zeros and thereby to some conservation laws is
however an open question.

\begin{figure}[t!]
\includegraphics[draft=false,clip=true,height=0.34\textwidth]{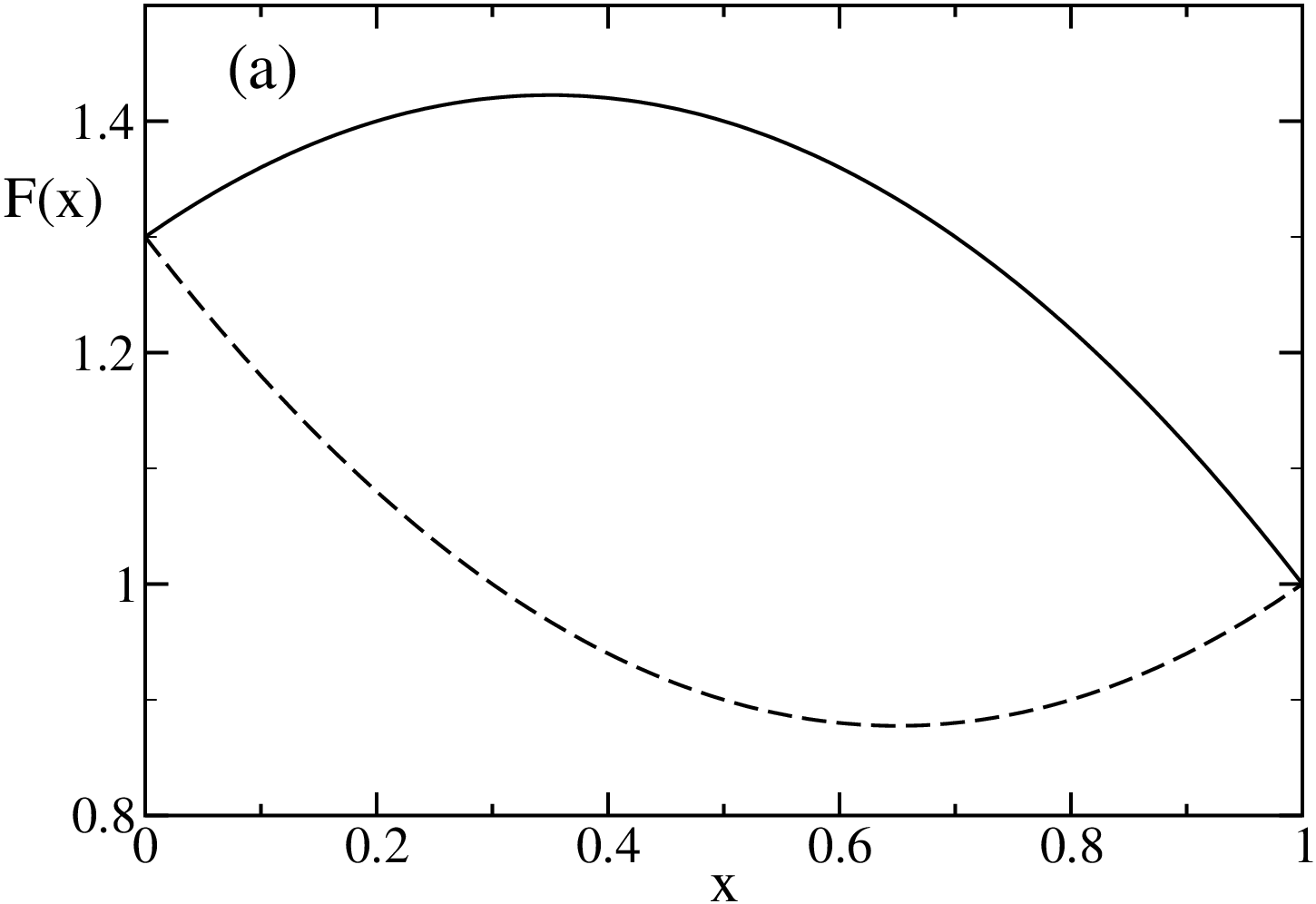}
\includegraphics[draft=false,clip=true,height=0.34\textwidth]{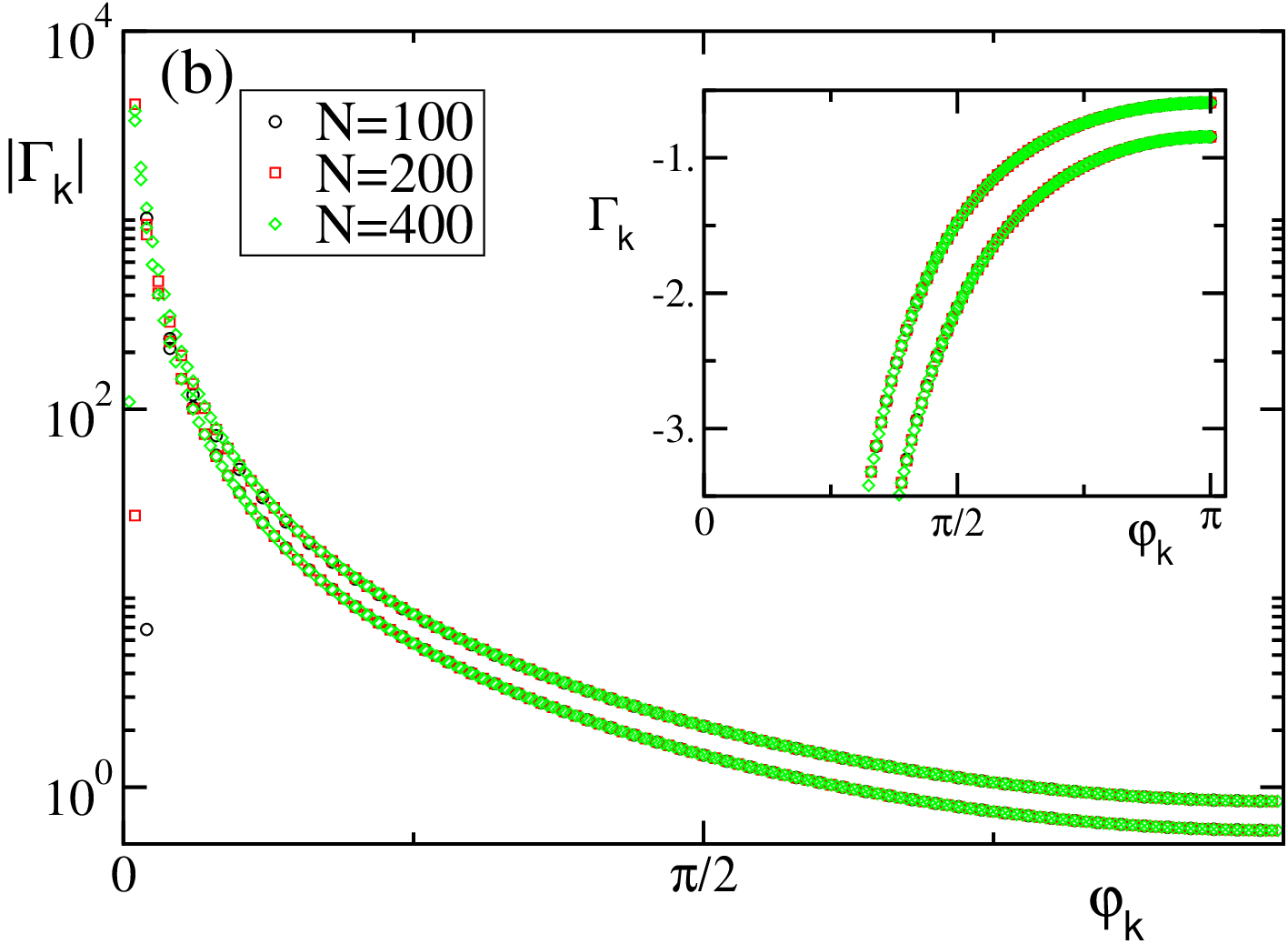}
\caption{
(Color online) (a) The two considered velocity fields: upper and lower curves
correspond to: $F(x)= 1.3 + 0.7 x -x^2$ (solid line) and $F(x)= 1.3 -1.3x + x^2$
(dashed line); (b) the lower (resp. upper) Floquet spectra are associated to the
lower (resp. upper) velocity field in Fig. \protect\ref{fig4}a, please notice
that in the inset due to the negative values of the spectra the correspondence
is reversed. The values of the other parameters are $g=0.4$, and $\alpha=6.0$.
}
\label{fig4}
\end{figure}

The choice of a periodic function $F(x)$ such as $a- \sin(2\pi x)$ is natural in
the context of coupled rotators, where $x$ is a true  phase and the 0, 1
values can be identified with one another as they correspond to angles
differing by $2\pi$. In the LIF model, 0 and 1 correspond to two different
membrane potentials (actually, the minimum and maximum accessible values) 
and there is no reason a priori to expect $F(1)=F(0)$, namely $\Delta F = F(1)-F(0)=-1$.
It is therefore important to understand
whether the different scaling behaviour is to be attributed to the presence of
a ``discontinuity" in the velocity field and if the sign of the difference
$F(1)-F(0)$ matters or not.  In order to clarify this point we
have investigated two parabolic fields with opposite concavities, but
identical (and negative) nonzero value of the velocity difference
at the extrema of the definition interval, i.e.  $\Delta F = -0.3$ 
(see Fig.~\ref{fig4}a for their graphical representation). 
The results plotted in
Fig.~\ref{fig4}b indicate that the presence of nonlinearities in the velocity
field do neither affect the scaling of the spectrum, that is still proportional
to $\tau^2$, nor the overall stability: the whole branch is strictly negative.

\begin{figure}[t!]
\includegraphics[draft=false,clip=true,height=0.34\textwidth]{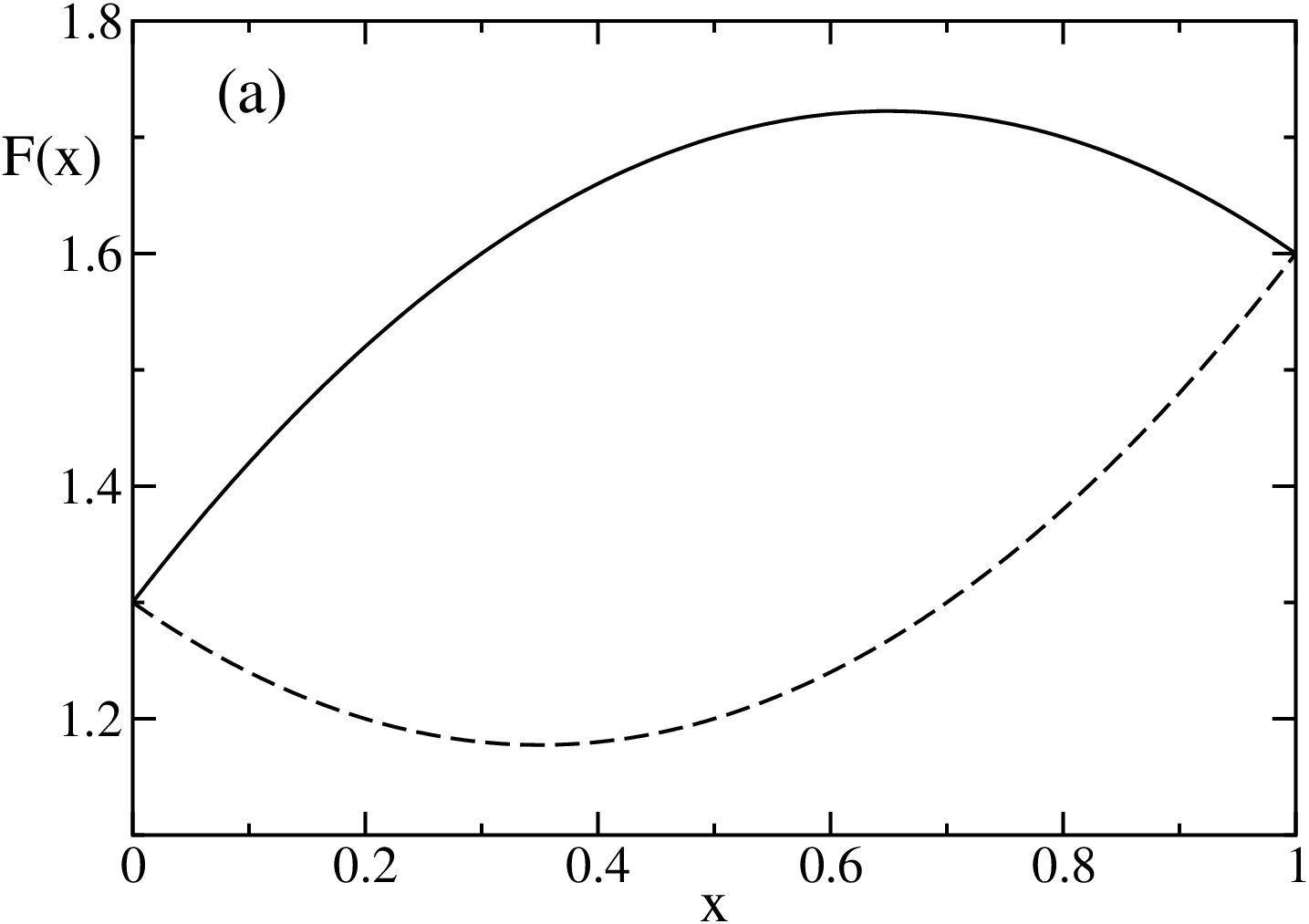}
\includegraphics[draft=false,clip=true,height=0.34\textwidth]{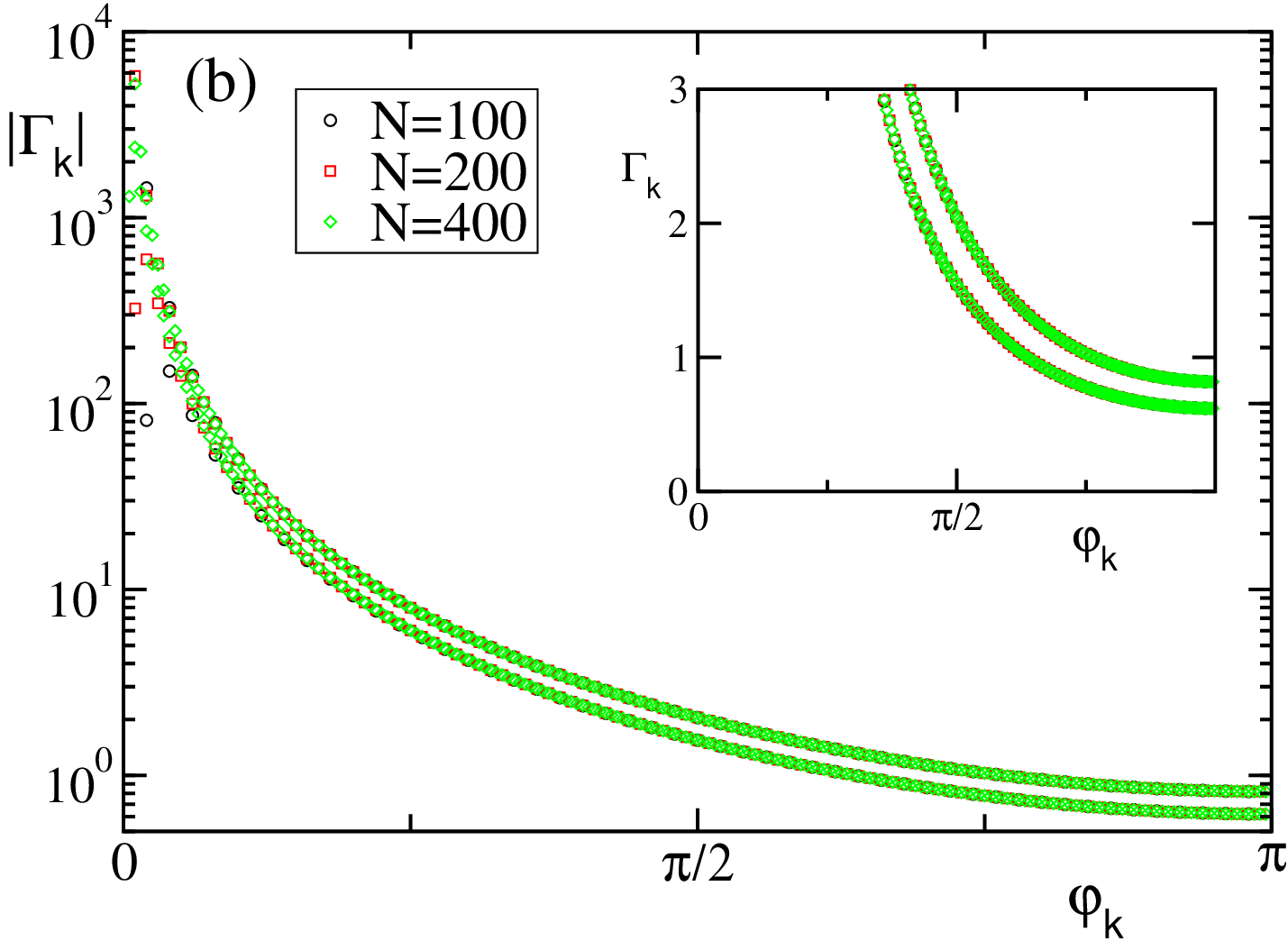}
\caption{
(Color online) (a) The two considered velocity fields: upper and lower curves
correspond to: $F(x)= 1.3 +1.3x - x^2$ and $F(x)= 1.3 - 0.7 x +x^2$; (b) the
lower (resp. upper) Floquet spectra are associated to the lower (resp. upper)
velocity field in Fig. \protect\ref{fig5}a.
The values of the other parameters are $g=0.4$, and $\alpha=6.0$.
}
\label{fig5}
\end{figure}

As a next step, we have investigated two increasing parabolic velocity fields
with opposite concavities, but identical and positive difference at the extrema
$\Delta F = 0.3$ (see Fig.~\ref{fig5}a for a graphical representation). The
results plotted in Fig.~\ref{fig5}b confirm once
again the $\tau^2$ scaling. However, the stability has changed: now the SW
component is positive, indicating that the splay state is
weakly unstable. Altogether, we can summarize the results under the conjecture
that all discontinuous velocity fields (i.e., where $F(1)\ne F(0)$) are
characterized by exponents that scale as $\tau^2$. Moreover, the stability
depends on whether on the average the field increases or decreases.
The analysis of several other velocity fields has confirmed this conjecture.

\begin{figure}[t!]
\includegraphics[draft=false,clip=true,height=0.34\textwidth]{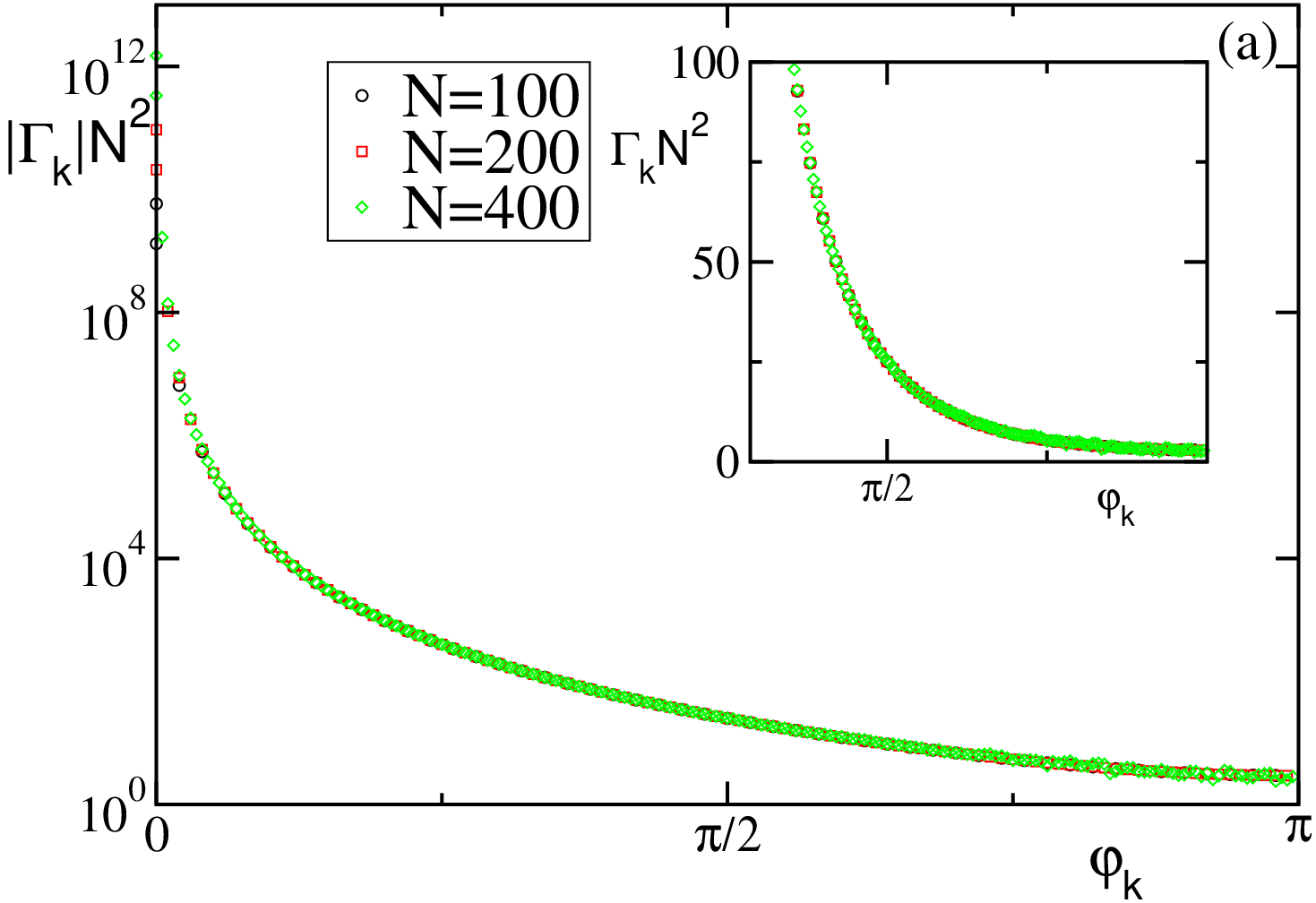}
\includegraphics[draft=false,clip=true,height=0.34\textwidth]{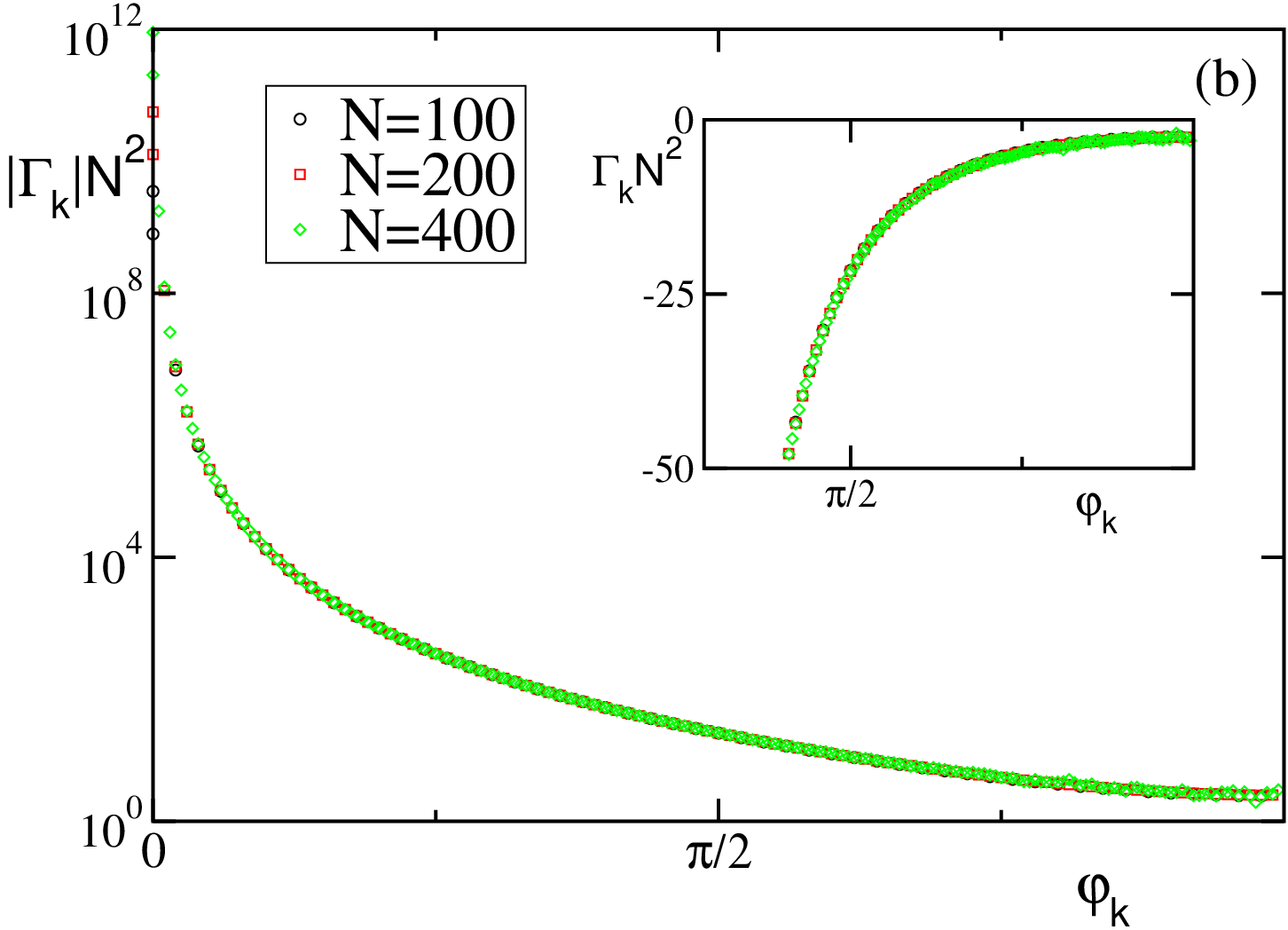}
\caption{
(Color online) Floquet spectra: (a) continuous parabolic field
$F(x) = 1.3 - x(x-1)$; (b) continuous sinusoidal field 
$F(x) = 1.3 - 0.25 \sin(\pi x)$ with period $2$. The data refer
to $g=0.4$ and $\alpha=6$ and have been obtained by employing an event-driven 
map including terms up to order $\cal{O}(1/N^5)$.
}
\label{fig6}
\end{figure}

In between the two classes of increasing and decreasing fields, there are
continuous fields. The neutral stability of the sinusoidal fields is logically
consistent with the observation that the stability depends on the sign of
$\Delta F$. In order to further explore the generality of the scenario, we have
analysed two other cases: (i) a parabolic field $F(x) = 1.3 - x(x-1)$; (ii) a
sinusoidal field $F(x) = 1.3 - 0.25 \sin(\pi x)$.
Both fields are ${\mathcal C}^{(0)}$, but not ${\mathcal C}^{(1)}$, since the
derivative in $x=1$ and $x=0$ differ from one another. In Fig.~\ref{fig6} we see
that the spectra scale as $1/N^4$.
This confirms that continuous functions exhibit an intermediate behaviour
between positive and negative discontinuities. The scaling behaviour, 
as $\tau^4$, has been verified by employing an event-driven map correct to 
order $\tau^5$.  Moreover, it is also interesting to notice the difference with respect to the
analytic sinusoidal functions. In fact, it seems that exactly zero exponents
are detected only for analytic velocity fields.

\begin{figure}[t!]
\includegraphics[draft=false,clip=true,height=0.3\textwidth]{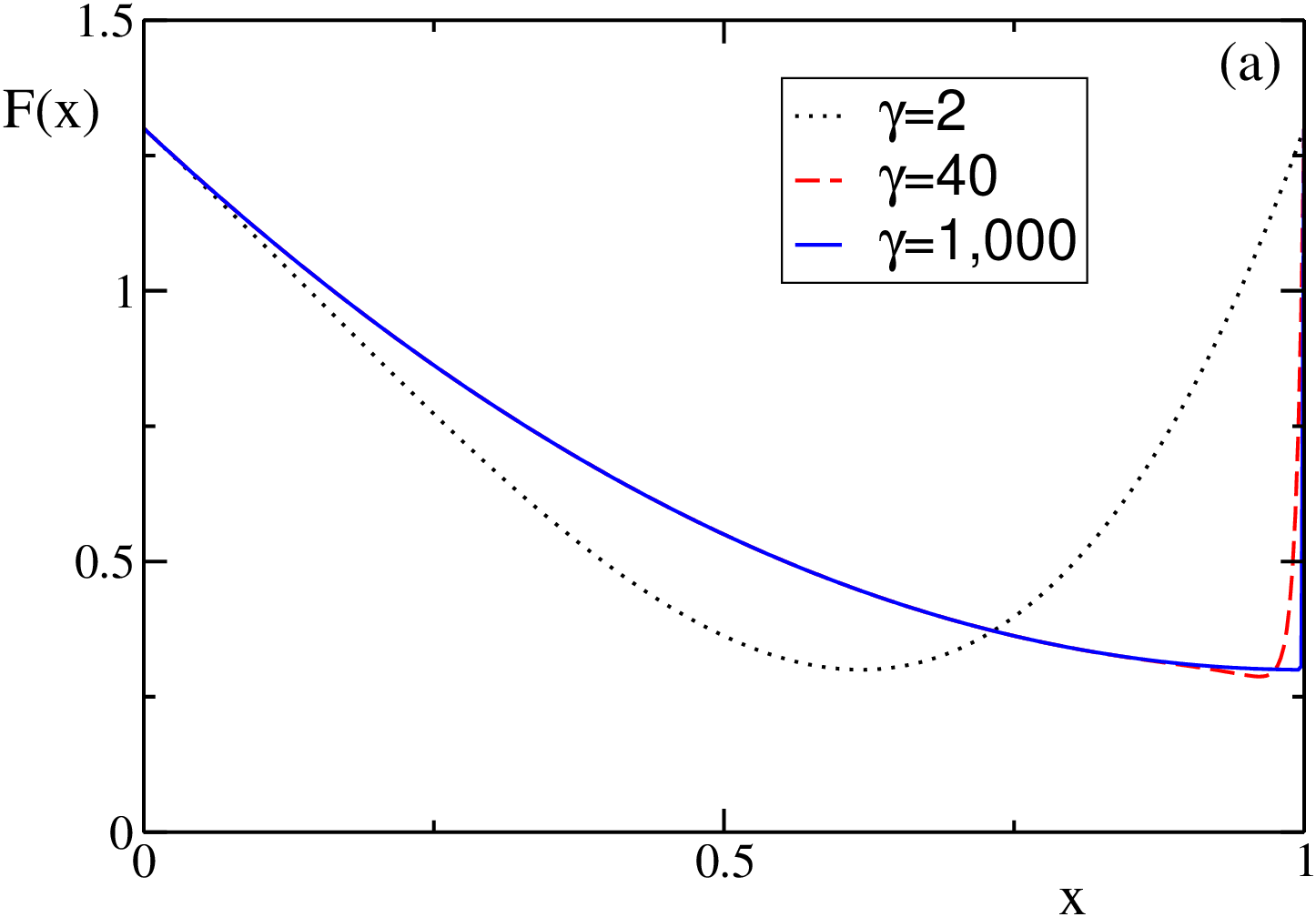}
\includegraphics[draft=false,clip=true,height=0.3\textwidth]{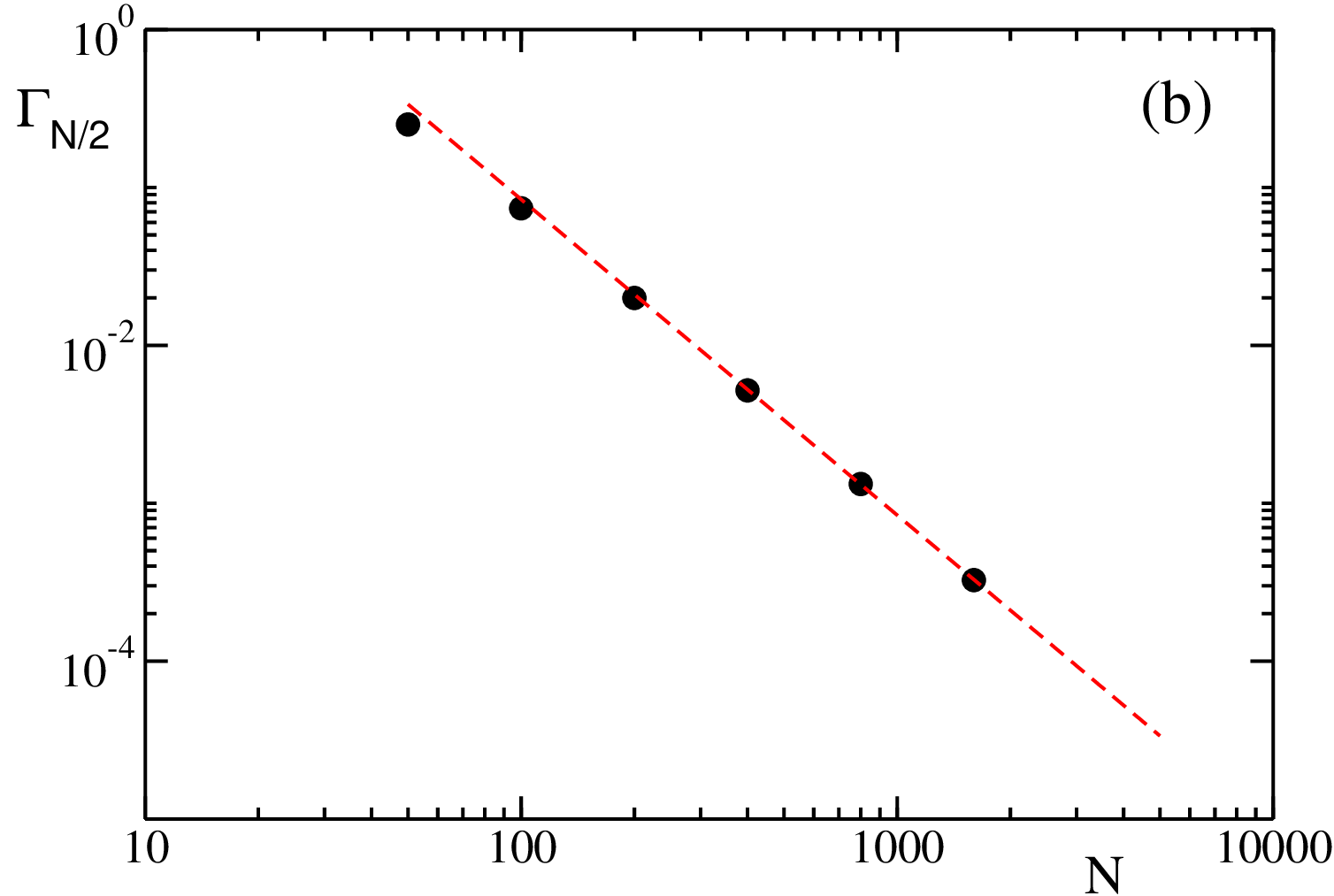}
\includegraphics[draft=false,clip=true,height=0.3\textwidth]{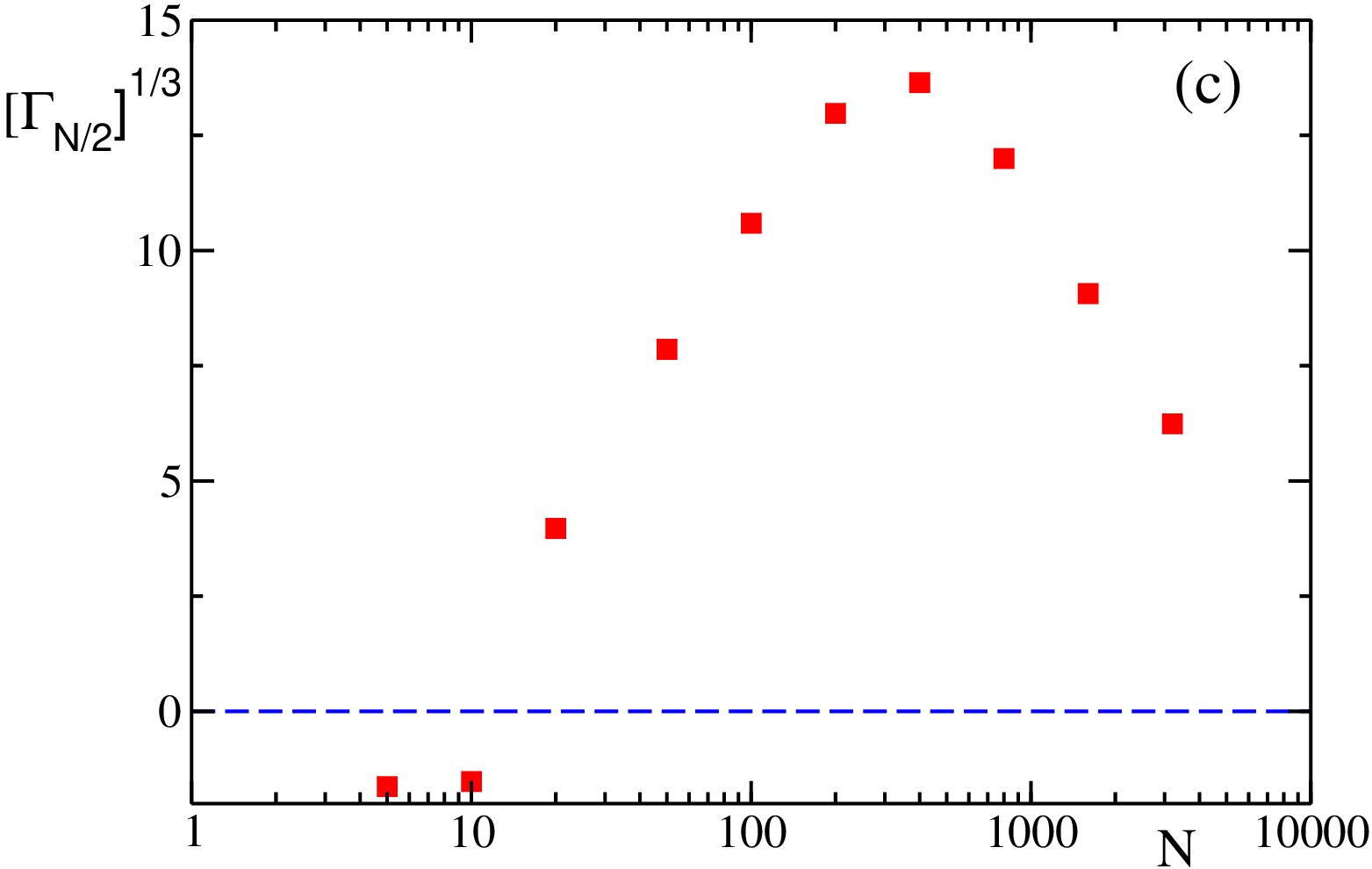}
\includegraphics[draft=false,clip=true,height=0.3\textwidth]{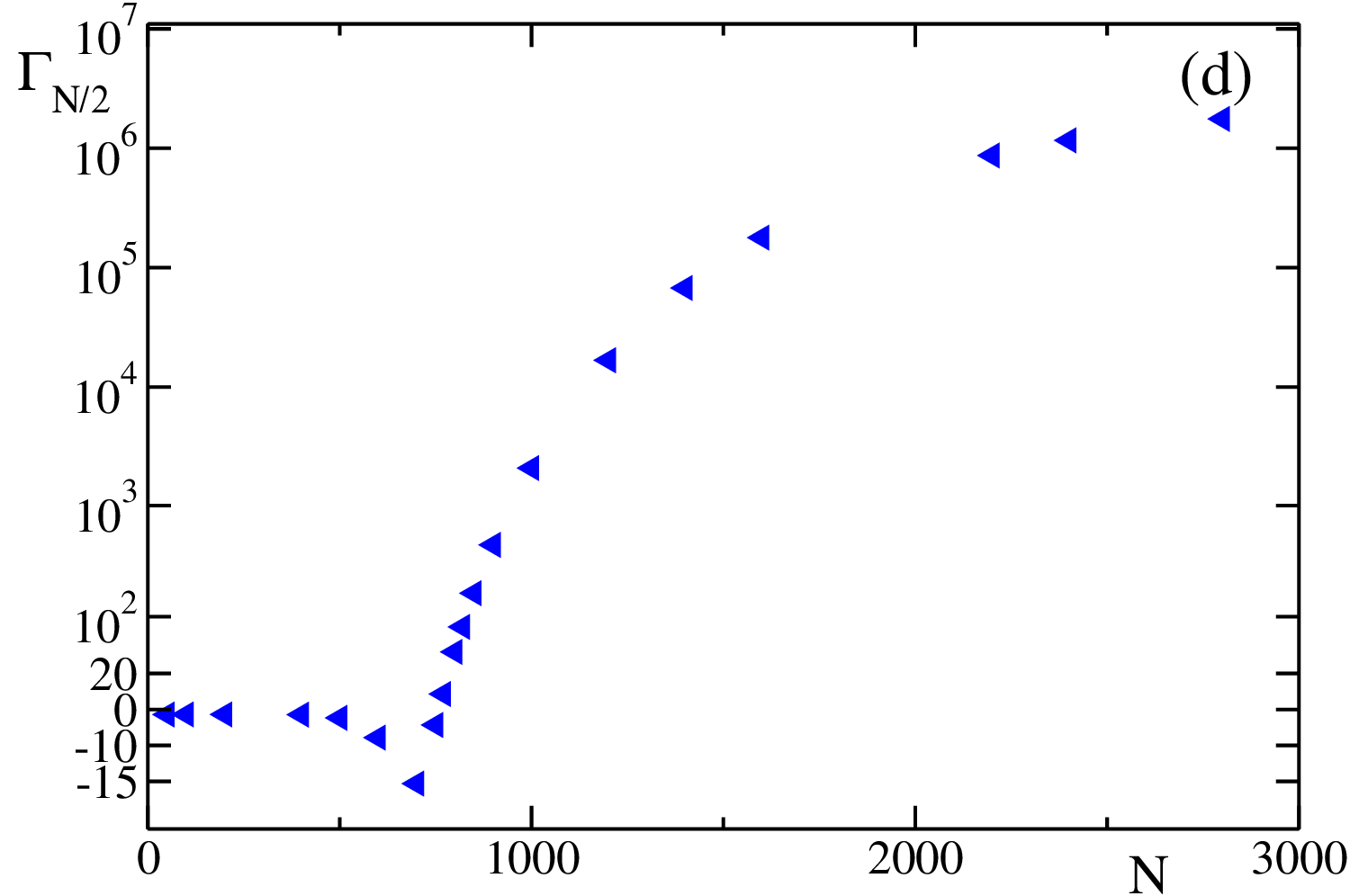}
\caption{
(Color online) (a) The velocity fields expressed by Eqs.~(\ref{nonlin_trasf})
for three different $\gamma$ values are reported in proximity of $x=1$.
The stability exponent $\Gamma_{N/2}$
corresponding to the highest frequency ($\varphi_{N/2} = \pi$) is plotted
in panels (b), (c), and (d) for the three different $\gamma$ values as a function of $N$.
In (b) , the (black) filled circles refer to $\gamma=2$ and 
the dashed (red) line represents a power-law decay $N^{-\beta}$ with
$\beta = 2$. In (c), $[\Gamma_{N/2}]^{1/3}$ is displayed for $\gamma = 100$;
the dashed (blue) line indicates the zero axis. Finally, for a better
visualization of the data for $\gamma = 10,000$, the vertical scale in (d) has
been obtained by first shifting $\Gamma_{N/2}$ by $20$ units and thereby using
exponentially separated units. All eigenvalues refer to $a=1.3$, $g=0.4$ and
$\alpha=6$.
}
\label{fig7}
\end{figure}

To further verify the scenario, we have introduced a velocity field
\bey
F(w)&=a+4w(w-1) 
\nonumber
\\
w&=(x^\gamma+x)/2
\label{nonlin_trasf}
\eey
parametrized by the exponent $\gamma$. The function is periodic but, upon
increasing $\gamma$, it becomes increasingly steeper in the vicinity of $x=1$
as shown in Fig.~\ref{fig7}(a). For $\gamma<10$ we observe the same
behaviour found for other periodic functions, i.e. the spectrum scales
rapidly to zero (like $1/N^4$). To exemplify this case let us
consider $\gamma =2$, as shown in Fig.~\ref{fig7}b the
eigenvalue $\Gamma_{N/2}$ decreases as $ \sim 1/N^2$ for sufficiently large $N$. 
Please notice that the eigenvalues are approaching zero from positive values in this case.
For $\gamma \sim 10^2$ the situation becomes more complicated:
for sufficiently small $N$-values $\Gamma_{N/2}$ is negative and almost constant
(indicating an $1/N^2$ scaling of the Floquet eigenvalues),
while by increasing $N$ it becomes positive and $\Gamma_{N/2} \to 0$
only for very large $N$, $\Gamma_{N/2} \to 0$ (see Fig.~\ref{fig7}c).
By increasing $\gamma$, the $1/N^2$ scaling region widens, but
the overall behaviour is maintained, as
shown in Fig.~\ref{fig7}d for $\gamma=10,000$.
This suggests that for not too large values of $N$, the system perceives the
field as if it was discontinuous, while at large $N$ it crosses to
continuous fields. This crossing is joined to a change of sign
from negative values (as expected for discontinous fields with $\Delta F < 0$)
to positive ones.

\begin{figure}[t!]
\includegraphics[draft=false,clip=true,height=0.34\textwidth]{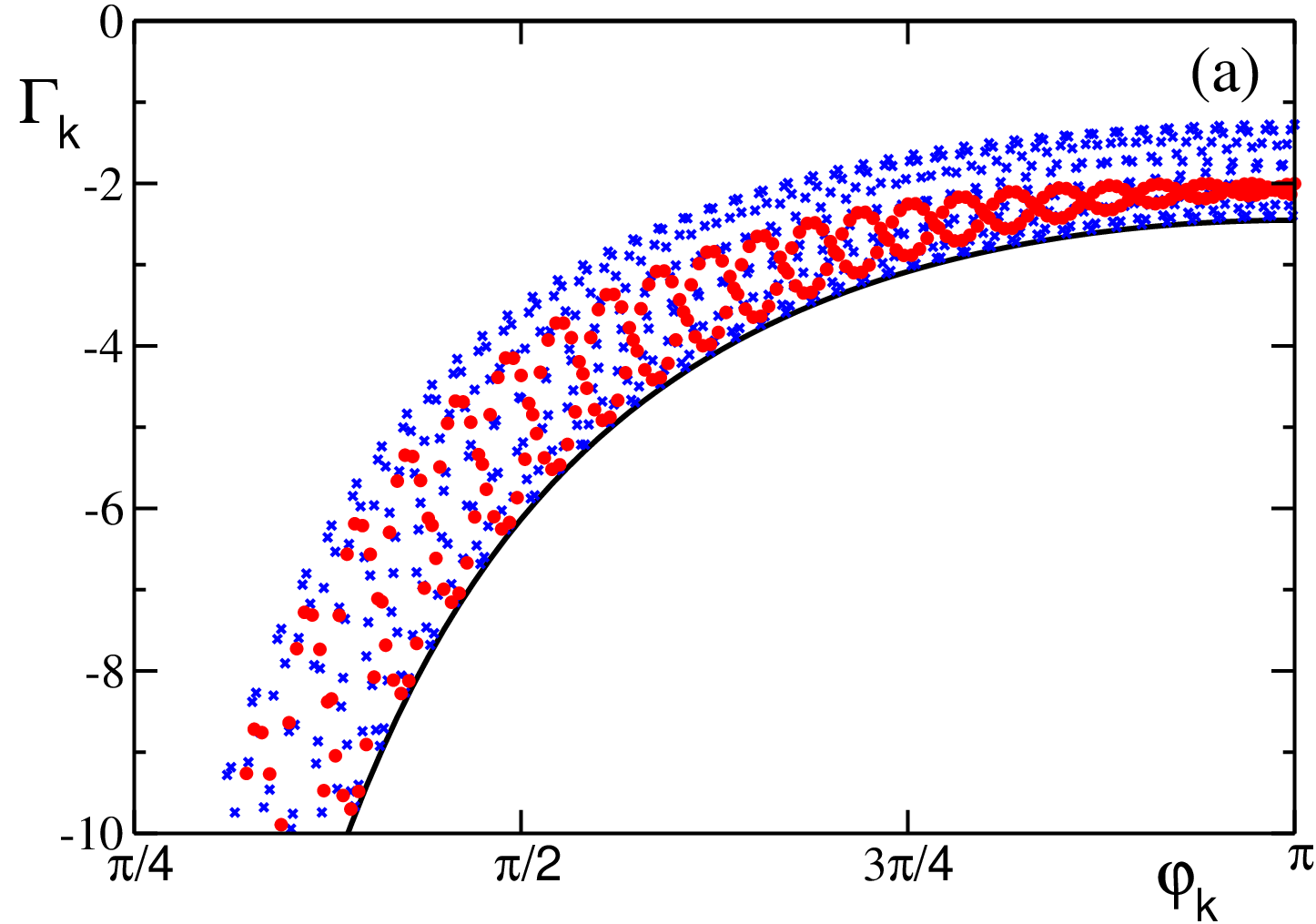}
\includegraphics[draft=false,clip=true,height=0.34\textwidth]{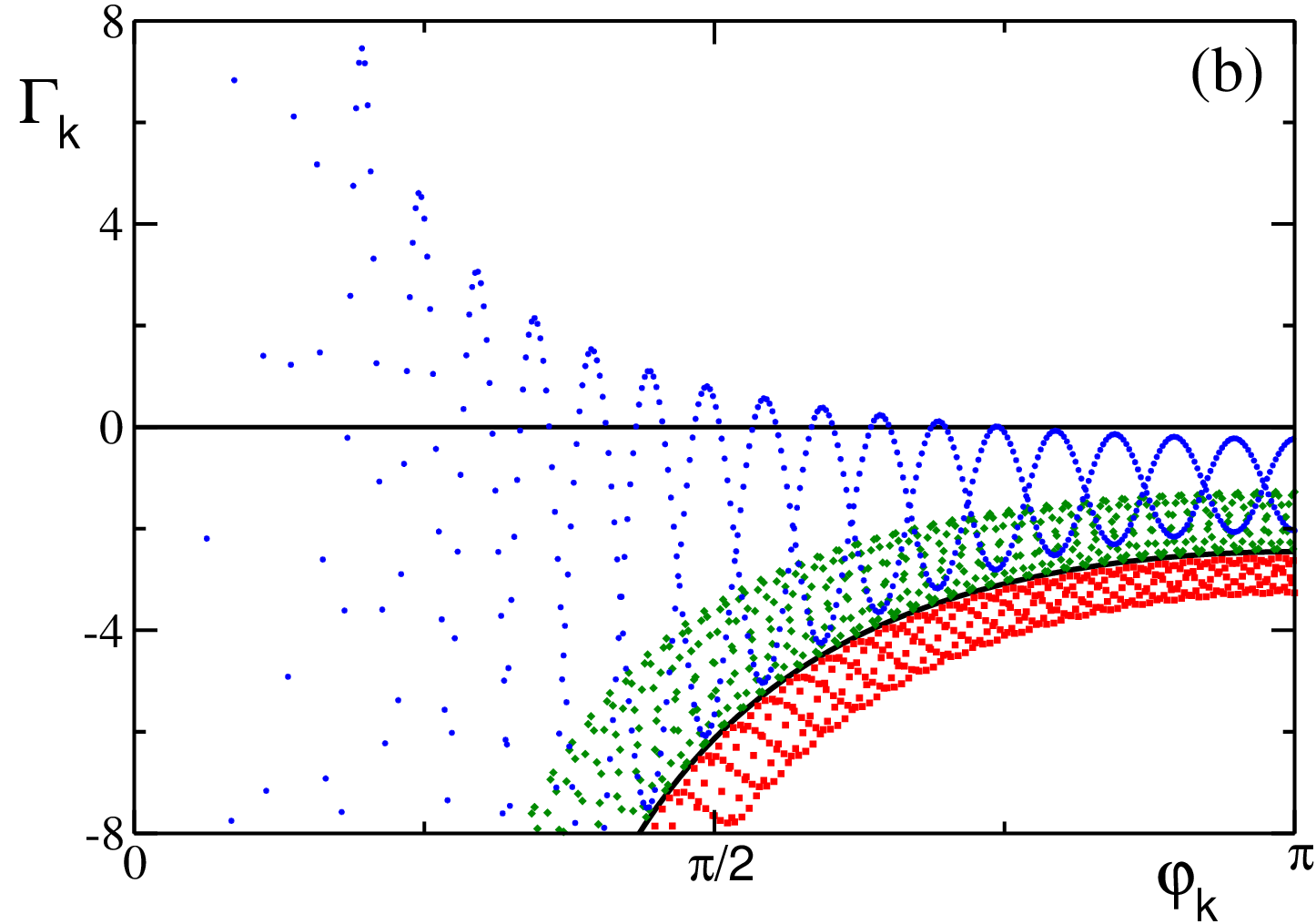}
\caption{
(Color online) Floquet spectra $\Gamma_k$ associated to the velocity
field (\ref{discontinuous}).
(a) Spectra for various $N$ with $b=+0.2$: $N=800$ (red) circles
and $N=1,600$ (blue) crosses. (b) Spectra for $N=1,600$ at different $b$-values:
namely, (red) squares refer to $b=-0.2$,  (green) diamonds
to $b=+0.2$ and (blue) stars to $b=+0.6$. For comparison also the Floquet
spectrum corresponding to a LIF neuron with the same parameters is reported
(black solid line). All the data refer to $a=1.7$, $g=0.4$ and $\alpha=6$.
}
\label{fig8}
\end{figure}

Finally, for the sake of mathematical generality, we have investigated the role
of an additional, intermediate, discontinuity in the velocity field.
More precisely, we have examined the piece-wise linear model,
\begin{equation}
F(x)= \left\{
\begin{array}{ll}
a -b/2 -x, & \mbox{for} \quad x \le 0.5 \\
a + b/2 -x, & \mbox{for} \quad x >  0.5
\end{array}
\right. 
\label{discontinuous}
\end{equation}
with a discontinuity of size $b$ at $x=0.5$. As shown in Fig.~\ref{fig8}a, as
a result of the discontinuity, the Floquet spectrum widens to cover a thick
band, that is filled in an increasingly uniform way, upon increasing $N$.
Depending whether the discontinuity is positive or negative, the band develops
towards higher or smaller values, respectively (see Fig.~\ref{fig8}b), while the
standard LIF spectrum (solid line) represents the locus of minima 
or maxima, respectively of such bands. Due to the finite thickness
of the band itself, unstable SW modes can appear even for $\Delta F < 0$.
Finally, for $b=1$, when $\Delta F= F(1)-F(0)=0$, the SW modes scale faster
than $1/N^2$ as for standard continuous functions. Altogether, the presence of
an additional discontinuity does not modify the scaling behaviour, that is
still controlled by the sign of $\Delta F$.

\section{Conclusions and open problems}
\label{five}
 
In this paper we have shown that the stability of splay states in pulse-coupled
oscillators with generic velocity fields $F(x)$ can be determined by rewriting
the dynamics as event-driven maps. In particular we focused our
attention on the SW modes. For discontinuous velocity fields,
like that associated to leaky integrate-and-fire neurons, we find that all SW
modes are stable, when the field on the average decreases
(i.e. $\Delta F<0$), and unstable in the opposite case. Notice that this weak
instability cannot be captured by the mean-field approach introduced
in \cite{abbott}, as the coarse-graining washes out SW modes. 
It is instructive to compare the role of $\Delta F$, with the results of
Refs.~\cite{mirollo,mauroy}. While studying an excitatory network of globally
coupled rotators (in the limit of $\delta$-like pulses), Mirollo and Strogatz
proved that the synchronous state is fully stable when $x(\phi)$ is concave-down
\cite{mirollo}, where the phase $\phi$ is nothing but the time variable (apart
from a scaling factor) and the evolution refers to the single neuron dynamics
$\dot x = F(x)$. Recently, these results have been complemented by the
observation that clustered states are stable (while the synchronous regime
is unstable), when $x(\phi)$ is concave-up~\cite{mauroy,notekirst}. 
It is easy to verify that in a
linear LIF neuron, the change of concavity is one-to-one connected with a
change of sign of $\Delta F$. More in general, a concave-up (-down) $x(t)$
implies that $\Delta F>0$ ($\Delta F<0$), while the opposite implication does
not hold. For instance, for $F(x) = 1.3 - \sin[\pi(9x-1)/4]$ ($x\in [0.1]$),
$x(t)$ exhibits even two changes of concavity and yet we find that the SW
modes are stable and scale as $1/N^2$, as expected, since
$\Delta F = - \sqrt{2}/2 $. Altogether, the condition arising from the sign of
$F$ is more general than that based on the sign of the concavity of $x(t)$,
but it refers to SW modes only.

Naively, one might think that that our results follow from the fact that
$\Delta F$ is the average derivative of the velocity field in the interval
$[0,1]$. If, on the average, $\dot x = (\Delta F) x$, it is natural to expect
exponential instability when $\Delta F>0$, stability in the opposite case, and
marginal stability for $\Delta F = 0$. However, to leading order, all SW modes
are ``marginally" stable. Presumably, there is some truth in the argument, but
some refinements are required to put it on a firm basis. Furthermore, the
$1/N^2$ scaling of the Floquet spectrum has been so far rigorously proved
for LIF neurons and has been confirmed by the numerical analysis of several
nonlinear fields. This includes the well known exponential integrate-and-fire
neurons (EIF) \cite{eif} which, having a discontinuous velocity field with
$\Delta F < 0$, are consistently characterized by stable SW modes.

The intermediate case of continuous velocity fields presents even more
subtleties. The are generically characterized by a $1/N^4$
scaling to zero, but we are unable to conclude whether the scaling law is
entirely determined by the analyticity properties of the velocity field.
Anyway, for fields composed of a few harmonics, it appears that all Floquet
exponents (with the exception of a finite number of them) are equal to zero.
These results suggest that perfectly marginal modes exist in a wide range of
cases than that those proved in \cite{watanabe} and shown in \cite{golomb_hansel}. 
The question is not of purely academic interest, since analytic velocity
fields are generically encountered when dealing with coupled rotators, where
$x$ is a true phase. Furthermore, the so-called quadratic integrate-and-fire
neurons (QIF) \cite{gerstner} belong to this class, as the velocity field 
is $F(x)=dx(1-x) +e$ and it is therefore continuous. We have verified that this
model is not an exception, as its short-wavelength spectrum scales faster
than $1/N^2$.

\acknowledgments
We acknowledge useful discussions with F. Ginelli, M. Timme, M. Wolfrum 
and S. Yanchuk, as well as a constructive interaction with S. Strogatz.
This work has been partly carried out with the support of the
EU project NEST-PATH-043309 and of the italian project ``Struttura e dinamica di reti
complesse'' N. 3001 within the CNR programme ``Ricerca spontanea a tema libero''. 


\end{document}